\documentclass[aps,prl,twocolumn,showpacs,superscriptaddress]{revtex4-1}  
\usepackage{amsmath}
\usepackage{amssymb}
\usepackage{graphicx}
\usepackage{epstopdf}

\begin{document}

\title{Two-Fluid Theory for Spin Superfluidity in Magnetic Insulators}

\author{B. Flebus}
\affiliation{Institute for Theoretical Physics and Center for Extreme Matter and Emergent Phenomena, Utrecht University, Leuvenlaan 4, 3584 CE Utrecht, The Netherlands}
\author{S. A. Bender}
\author{Y. Tserkovnyak}
\affiliation{Department of Physics and Astronomy, University of California, Los Angeles, California 90095, USA}
\author{ R. A. Duine}
\affiliation{Institute for Theoretical Physics and Center for Extreme Matter and Emergent Phenomena, Utrecht University, Leuvenlaan 4, 3584 CE Utrecht, The Netherlands}

\begin{abstract}
We investigate coupled spin and heat transport in easy-plane magnetic insulators. These materials display a continuous phase transition between normal and condensate states that is controlled by an external magnetic field. Using hydrodynamic equations supplemented by Gross-Pitaevski phenomenology and magnetoelectric circuit theory, we derive a two-fluid model to describe the dynamics of thermal and condensed magnons, and the appropriate boundary conditions in a hybrid normal-metal$|$magnetic-insulator$|$normal-metal  heterostructure. We discuss how the emergent spin superfluidity can be experimentally probed via a spin Seebeck effect measurement.
\end{abstract}



\maketitle

\textit{Introduction}.|It has been many years since Kapitza first observed that helium, when cooled  below a temperature of $2.17$ K, displays properties attributable to a new quantum phase of matter \cite{kapitz}, such as the ability to flow without  dissipation through thin capillaries, the quantization of the vorticity  and a record thermal conductivity.
These properties are well understood within the framework of the two-fluid model proposed independently by Tisza \cite{tisza} and Landau \cite{landau22}, in which He II is described as a mixture 
of a normal fluid, which is viscous and carries all the entropy of the system, and a superfluid that flows without  friction and carries no thermal energy.

Only a few years later, the two-fluid model successfully threw light upon the apparent absence of the usual thermoelectric effects, such as the Seebeck and the Peltier effects, in the superconducting state \cite{gin}. Indeed, in superconductors, all the conventional thermoelectric properties vanish due to the coexistence of the thermal quasiparticle current with a dissipationless supercurrent that counterflows with it. The analogy between the supercurrent of electric charge in superconductors and the mass superflow in helium stems from the underlying common origin of these phenomena, i.e., the spontaneous breaking of the $U(1)$ symmetry underlying Bose-Einstein condensation (BEC, of either atoms or Cooper pairs) and the associated macroscopic quantum coherence. Therefore, a superfluid phase can be described by a two-fluid model, in which the condensed and itinerant atoms are, loosely speaking, identified with the superfluid and normal components, respectively. This concept can be extended to a variety of systems exhibiting $U(1)$ symmetry breaking and thus the coexistence of a normal and a Bose-Einstein condensed fluids, such as excitons \cite{ex1,ex2}, polaritons  \cite{pol1,pol2},  and magnons \cite{mag1,mag2,mag3}.  

\begin{figure}[b]
\includegraphics[width=0.9\linewidth]{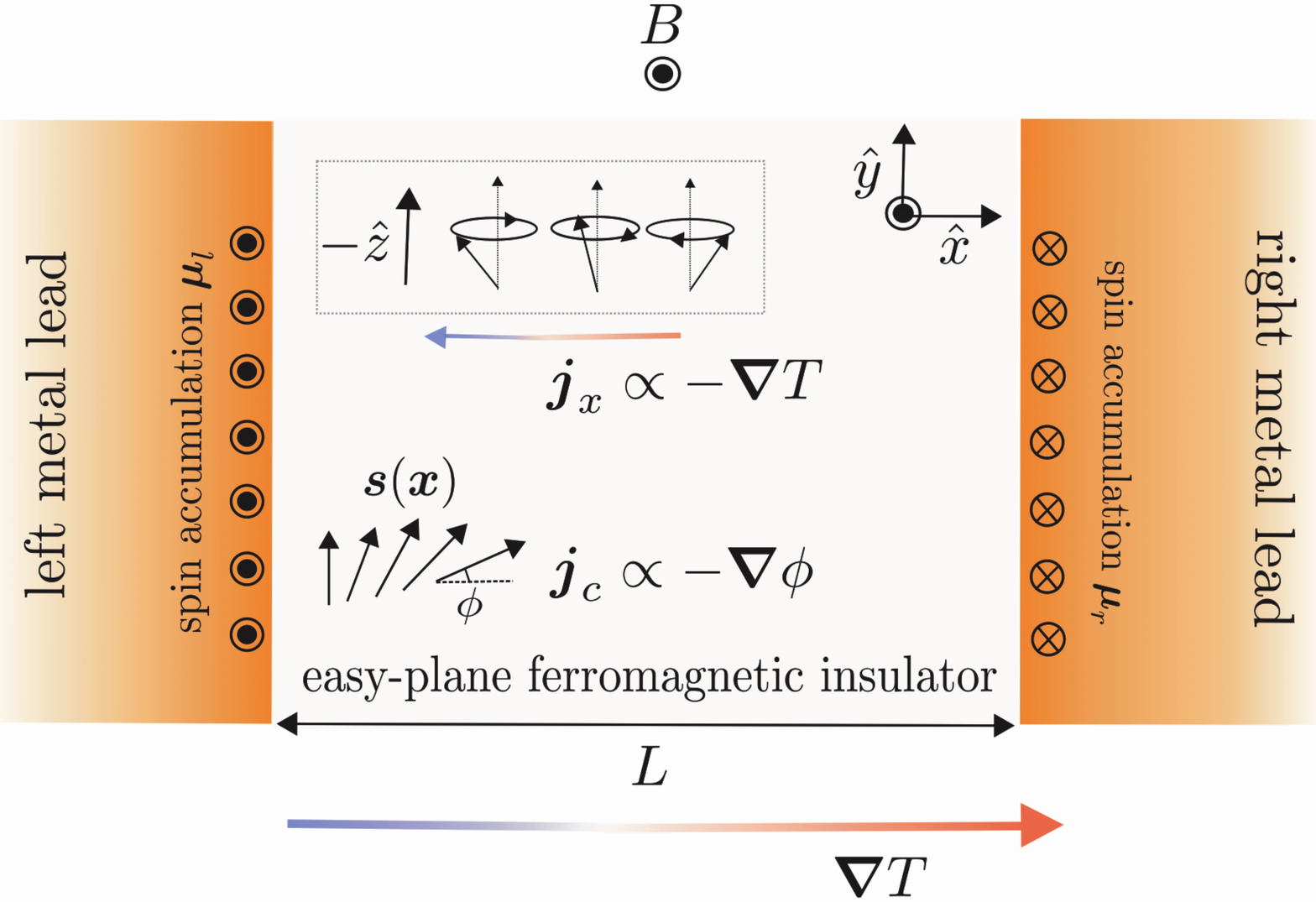}
\caption{Normal-metal$|$easy-plane insulator$|$normal-metal hybrid heterostructure. The state of the equilibrium magnetization, which is determined by the interplay between the magnetic field $B$ and the anisotropy energy $K$, can be perturbed by magnon transport driven by temperature gradient $\boldsymbol{\nabla}T$ and spin accumulations $\boldsymbol{\mu}_{l,r}=\mu_{l,r}\mathbf{\hat{z}}$ sustained by the metal leads. At low magnetic fields, the spin Seebeck current $\textbf{j}_{x}$ induced by the temperature gradient $\boldsymbol{\nabla}T$ coexists with a superfluid spin counterflow $\textbf{j}_{c}$, as discussed in the text.}
\label{scheme}
\end{figure}

A growing interest has recently arisen  
in magnonic systems as promising setups for achieving room-temperature 
Bose-Einstein condensation, motivated in part by the experimental progress of 
Demokritov~\textit{et al.} \cite{dem} on parametrically pumped magnon condensates. More recently, a theoretical proposal for the realization of a BEC of  magnons by means of direct spin current injection from an adjacent normal metal with strong spin-orbit coupling was put forward by Bender~\textit{et al.} \cite{bender}. Unlike BEC of real particles, BEC of quasiparticles and, in particular, quasiequilibrium  magnons does not require low temperatures, since the high densities of magnons needed for the condensate to form can be produced via external pumping or by tuning the magnetic field, which is facilitated by their small effective mass (corresponding to strong exchange). 
In this Letter, we focus on a ferromagnetic insulator with easy-plane magnetic anisotropy as a simple model system that displays a transition between normal and BEC phases and exhibits superfluid behavior. The magnet is sandwiched between two metallic reservoirs that act like thermal baths, set at two different temperatures, and that may provide spin accumulation via the spin Hall effect (as illustrated in Fig.~\ref{scheme}).  The temperature difference applied across the ferromagnet induces a spin current into normal metals, which can be measured as an inverse spin Hall voltage and is dubbed the spin Seebeck effect \cite{experiment}.   By sweeping the magnetic field in the $z$ direction, the system can be tuned to a state where the ($xy$) easy-plane rotational symmetry is spontaneously broken, and which, as a result, supports collective spin currents. We show that the spin Seebeck effect is then diminished, as a result of counterflow between condensate and thermal spin currents. As a practical utility, our results may provide novel routes to control thermal spin currents. 

\begin{figure}[t]
\includegraphics[width=0.9\linewidth]{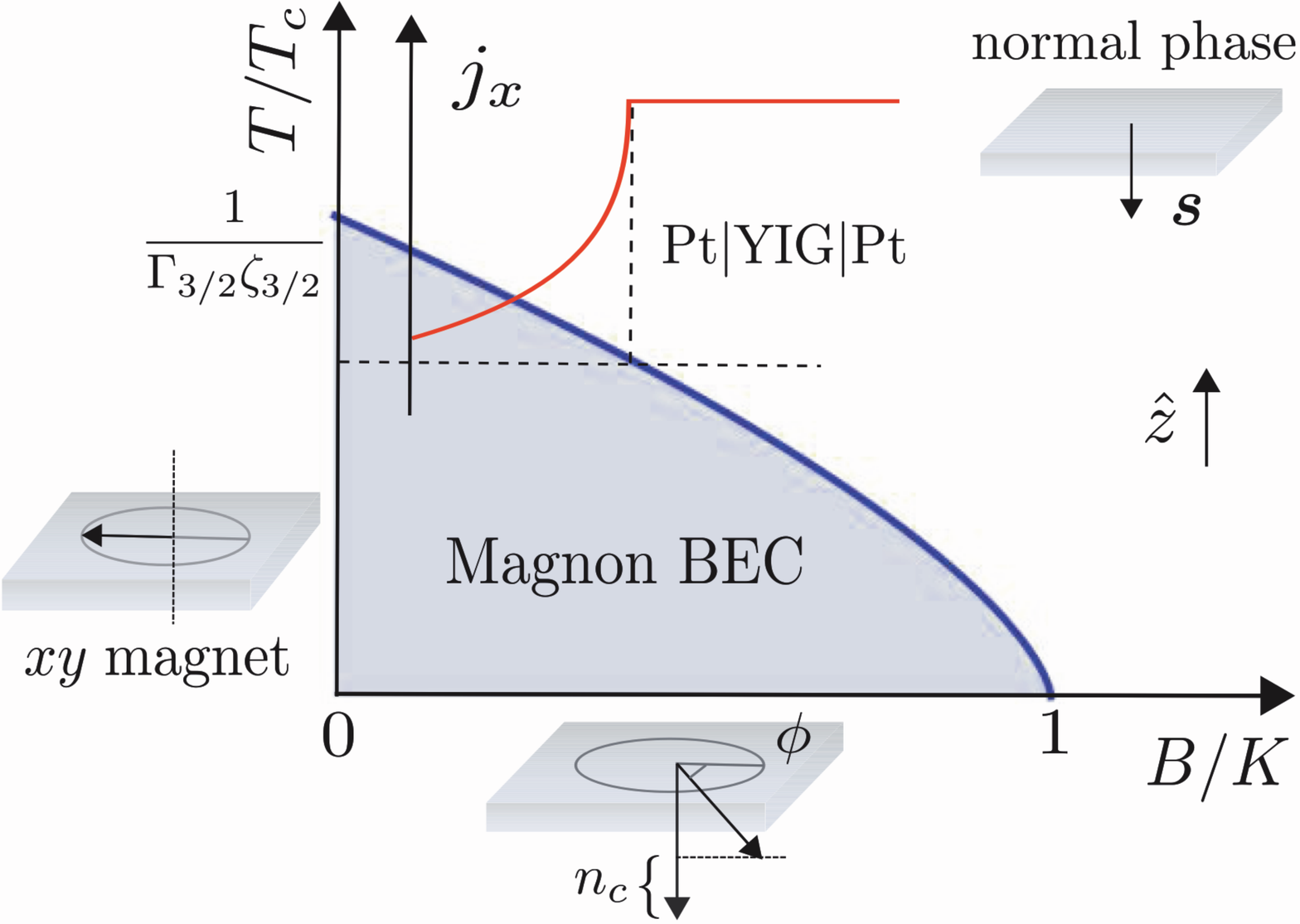}
\caption{Equilibrium phase diagram. The condensate phase boundary is at $T/T_c = (1-B/K)^{2/3}/\Gamma_{3/2} \zeta_{3/2}$, where $T_{c}\equiv A s^{2/3}$ estimates the Curie temperature. In the normal phase, the net spin density $\mathbf{s}$ is oriented along the (negative) $z$ axis; the condensate spontaneously breaks U(1) symmetry around the $z$ axis, as manifested by a static canting of the magnetization, whose deviation from its normal-state equilibrium value along the $z$ axis is parametrized by the condensate density $n_{c}$. In the absence of an applied field $B$, the ferromagnet is a planar $xy$ magnet. The reduction of the spin Seebeck current $j_{x}$ (red curve) as the magnetic field $B$ decreases below the transition point, at a fixed $T$,  is a direct and observable signature of superfluidity.}
\label{phasediagram}
\end{figure}

\textit{Model and hydrodynamic equations}.|We consider the following model Hamiltonian for an easy-plane magnetic insulator subjected to a field $B$ oriented along the $z$ axis:
\begin{equation}
\mathcal{H}=\int d^{3}r \left(- \frac{A}{2s} \hat{\mathbf{s}} \cdot \nabla^{2} \hat{\mathbf{s}} + B\hat{s}_{z} + \frac{K}{2s}\hat{s}^{2}_{z}\right),
\label{discreteH}
\end{equation}
where $\hat{\mathbf{s}}$ is the spin density operator (in units of $\hbar$), $A$ the exchange stiffness, $K>0$ the constant governing the strength of the local easy-plane anisotropy, and $s$ the saturation spin density. Performing the Holstein-Primakoff transformation \cite{holstein}, $\hat{s}_{z}=\hat{\Phi}^{\dagger} \hat{\Phi}-s$ and $\hat{s}_{-}=\sqrt{2s-\hat{\Phi}^{\dagger} \hat{\Phi}} \hat{\Phi}$, it is straightforward to recast the Heisenberg dynamics of $\hat{\mathbf{s}}$ as a superfluid coupled to a normal cloud (see, e.g., Ref.~\cite{benderduine}). By, furthermore, including phenomenologically the Gilbert damping constant $\alpha$, the corresponding Gross-Pitaevksi equation (following the Popov approximation \cite{gp}) reads as
\begin{align}
(i-\alpha)\hbar \partial_{t} \Phi=\left( \hbar\Omega + Kn_{c}/s - i R \right) \Phi - A\nabla^2 \Phi.
\label{stocastic}
\end{align}
Here $\Phi\equiv\langle \hat{\Phi} \rangle=\sqrt{n_{c}} e^{-i \phi}$ is 
the superfluid order parameter, with $\phi$ being the 
precessional angle of the magnetization density in the \textit{xy} plane and $n_c$ ($n_x$) condensed (normal) magnon density. In particular, $s_z=n_c+n_x-s$. We are assuming small deviations from the ground state (in the absence of anisotropy), so that $n_c+n_x\ll s$, throughout. $\hbar 
\Omega\equiv B-K(1-2n_x/s)$ is the normal-phase magnon gap, and the collisional term $R$ 
describes the coupling to the finite-temperature normal cloud \cite{stoof}, which is defined by 
$\hat{\phi}\equiv\hat{\Phi}-\Phi$, with $\langle \hat{\phi}
^{\dagger} \hat{\phi} \rangle$ being the normal cloud density $n_x$. At 
zero temperature (and thus $R\to0$), Eq.~(\ref{stocastic}) recasts the Landau-Lifshitz-Gilbert equation \cite{landau} for
small-angle dynamics of the spin density around the $-\mathbf{z}$ direction (see Fig.~\ref{phasediagram}). It is, furthermore, illuminating to rewrite Eq.~(\ref{stocastic}) as the superfluid hydrodynamic equations:
\begin{subequations}\begin{align}
\dot{n}_{c} + \boldsymbol{\nabla} \cdot \mathbf{j}_{c} &= - \Gamma_{cx} - 2 \alpha \omega n_c,  \label{hydrofirst} \\
\hbar \omega &=  \hbar \Omega+K\frac{n_{c}}{s}+A\left[ \left( \boldsymbol{\nabla} \phi \right)^{2}-\frac{\nabla^2\sqrt{n_c}}{\sqrt{n_c}}\right],
\label{hydrosecond}
\end{align}\label{hydro}\end{subequations}
where $\omega=\dot{\phi}$ is the condensate frequency and  $\mathbf{j}_{c}=n_{c} \mathbf{v}_{c}$ the condensate spin current, where $\mathbf{v}_{c}=-\hbar\boldsymbol{\nabla}\phi/m$ and $m\equiv\hbar^{2}/2A$ is the kinetic magnon mass. $\Gamma_{cx}= 2n_{c} R/\hbar$ is the collision term describing equilibration between the condensate and the thermal cloud, defined as $\Gamma_{cx}=n_{c}(\hbar\omega-\mu)/\tau_{cx}T$ \cite{gp}, with $\tau_{cx}$ parametrizing the collision time between condensed and thermal magnons \cite{suppl1}. Chemical potential $\mu$ and temperature $T$ parametrize the Bose-Einstein distribution of the thermal cloud.

The equilibrium phase diagram of the easy-plane condensate is shown in Fig.~\ref{phasediagram}, which is obtained by a mean-field self-consistency analysis for $n_c\geq0$ coupled to the thermal cloud \cite{suppl1}. In the following, we will be interested in the linear response of magnons to a temperature gradient. Linearizing with respect to small nonequilibrium variables|$\omega$, $\mathbf{v}_{c}$, and $\delta n_c\equiv n_c-n_c^{(0)}$ for the condensate and $\mu$ and $\delta T\equiv T-T^{(0)}$ for the cloud|Eqs.~(\ref{hydro}) become
\begin{subequations}\begin{align}
\delta \dot{n}_{c} + n_c\boldsymbol{\nabla}\cdot\mathbf{v}_c=& \frac{n_{c}}{\tau_{cx} }\frac{\mu - \hbar\omega}{T} -2\alpha\omega n_c,\label{lha}\\
\hbar \omega=& K \frac{\delta n_{c}+2\delta n_{x}}{s}-A\frac{\nabla^2\delta n_c}{2n_c}. \label{lhb}
\end{align}\label{lh}\end{subequations}
Here $\delta n_x\equiv n_x-n_x^{(0)}$ can be expanded in terms of $\mu$ and $\delta T$ (disregarding its subleading dependence on $\delta n_c$). The superscript $(0)$, which was dropped in Eqs.~\eqref{lh} without danger of ambiguity, denotes the corresponding equilibrium values in the absence of the thermal flux.

The above condensate equations are complemented by hydrodynamic equations for the thermal cloud, which can be easily constructed within the Boltzmann transport theory \cite{suppl1}:
\begin{subequations}\begin{align}
\delta \dot{n}_x + \boldsymbol{\nabla} \cdot \mathbf{j}_x&= \frac{n_{c}}{\tau_{cx}}\frac{\hbar\omega-\mu}{T} - g_{n \mu} \mu - g_{n T} ( T - T_p) , \label{nua}\\
\delta \dot{u} + \boldsymbol{\nabla} \cdot \mathbf{j}_q&= - g_{u T} ( T - T_{p}) - g_{u \mu} \mu.
\label{nub}
\end{align}\label{nu}\end{subequations}
Here $u$ is the energy density of the thermal cloud, $T_p$ is the phonon temperature, and the $g$ coefficients parametrize relaxation of magnons by the (phononic) environment. [Note that a contribution to the energy rate equation \eqref{nub} from the condensate-cloud 
scattering is missing as it is quadratic in the nonequilibrium bias: $
\delta \dot{u}|_{cx} \propto \hbar \omega(\hbar \omega-\mu).$] The linear response spin, $\mathbf{j}_x$, and heat, $\mathbf{j}_q$, current densities, furthermore, can be expanded
as
\begin{equation}
\mathbf{j}_x =-\sigma \boldsymbol{\nabla} \mu - \varsigma \boldsymbol{\nabla} T,~~~\mathbf{j}_q = - \kappa \boldsymbol{\nabla} T-\rho \boldsymbol{\nabla} \mu,
\label{currents}
\end{equation}
where $\sigma$, $\kappa$, $ \varsigma $, and $\rho$ are respectively the bulk spin and heat conductivities and the intrinsic spin Seebeck and Peltier coefficients.

\textit{Boundary conditions}.|The spin and heat flow across the sample must be determined consistently with the boundary 
conditions defined at the F$|$N interfaces at $x=0, L$.
 Accounting for interfacial static spin-transfer and spin-pumping 
 torques, the linearized $z$ component of the condensate spin 
 current density injected from the left reservoir with a nonequilibrium spin 
 accumulation $\boldsymbol{\mu}_l= 
 \mu_l\mathbf{z} $  is given by  \cite{magneto}
\begin{equation}
 j_{c}|_{x=0}=n_c g^{\uparrow \downarrow}_l (\mu_l - \hbar\omega)/ 2\pi\hbar s ,
 \label{BCcondensate}
\end{equation}
where $g_l^{\uparrow \downarrow}$ is the real part of the (dimensionless) spin mixing conductance (per unit area).
The thermal spin and heat currents flowing across the left interface are given by
\begin{subequations}\begin{align}
j_x|_{x=0} &= G  (\mu_l - \mu)|_{x=0} + S  ( T_l - T)|_{x=0}  , \\
j_q|_{x=0}  &= \textit{K}  ( T_l - T)|_{x=0} + \Pi ( \mu_l - \mu )|_{x=0}  , 
\end{align}\label{BBC}\end{subequations}
Here $T_l$ is the electron temperature and $G$, $K$, $S$, and $\Pi$ are the interfacial magnon spin and thermal conductances and spin Seebeck and Peltier coefficients, respectively. 

The boundary conditions, Eq.~(\ref{BCcondensate}) and Eq.~(\ref{BBC}) along with the analogous expressions for the right interface, together with the two-fluid hydrodynamic relations, Eqs.~(\ref{lh}) and (\ref{nu}), constitute a complete set of linearized equations from which we can yield solutions for all the dynamical variables. We will now solve this problem in a steady state (i.e., $\delta\dot{n}_c=\delta\dot{n}_x=\delta\dot{u}=0$ and $\omega={\rm const}$), when the normal-metal reservoirs are thermally biased: $T_l=T-\Delta T/2$ and $T_r=T+\Delta T/2$. We will suppose, for simplicity, that the phononic heat transport and thermal profile are only weakly disturbed by the magnons, so that $T_p=T+\Delta T(x/L-1/2)$, where we, furthermore, neglected interfacial Kapitza resistances.

\textit{Results}.|Let us investigate the flow of magnonic spin and heat across a mirror-symmetric N$|$F$|$N structure driven by a small temperature bias $\Delta T$. We will consider two limiting cases: the magnet is sandwiched (1) between two heavy metals acting as good spin sinks (as may be exemplified by Pt$|$YIG$|$Pt), in which case $\mu_{l,r}=0$, or (2) between two light metals being perfectly poor spin sinks (possibly approximated by Cu$|$YIG$|$Cu), in which case spin accumulations build in each lead to block the total spin current across the interfaces, $j_c+j_x\to0$ at $x\to0,L$.

Since the spin-preserving relaxation of magnon distribution towards the phonon temperature, as parametrized by $g_{uT}$ in Eq.~\eqref{nub}, does not rely on relativistic spin-orbit interactions, we may expect it to be an efficient process at high temperatures (stemming, e.g., from the modulation of exchange coupling by lattice vibrations). The corresponding lengthscale, which is governed by the inelastic magnon-phonon scattering, $\lambda_u\equiv\sqrt{\kappa/g_{uT}}$, can therefore be taken to be shorter than other relevant lengthscales, which are associated with relativistic physics (i.e., $\lambda_n$ and $\lambda_{cx}$ defined below). In this regime, we can set $T\to T_p$, which decouples the spin transport from heat dynamics, resulting, in the steady state, in the following diffusion equation for magnons:
\begin{equation}
\partial_x^2\mu-(\mu-\hbar\omega)/\lambda^2_{cx}-\mu/\lambda_n^2=0,
\end{equation}
which is solved by
\begin{equation}
\mu=(\lambda_{m}/\lambda_{cx})^2\hbar\omega+c_le^{-x/\lambda_m}+c_re^{(x-L)/\lambda_m}.
\end{equation}
Here $\lambda^{-2}_m\equiv\lambda^{-2}_n+\lambda^{-2}_{cx}$, $\lambda_n\equiv\sqrt{\sigma/g_{n\mu}}$ is the thermal magnon diffusion length, and $\lambda_{cx}\equiv\sqrt{\sigma\tau_{cx} T/n_c}$ is the condensate-cloud equilibration length (where $n_c$ is the condensate equilibrium density according to the phase diagram in Fig.~\ref{phasediagram}). The boundary conditions are given by
\begin{subequations}\begin{align}
j_x(0)&=G_*c_l- \varsigma \Delta T/L=G[\mu_l-\mu(0)],\\
j_x(L)&=-G_*c_r- \varsigma \Delta T/L=G[\mu(L)-\mu_r],
\end{align}\label{bcx}\end{subequations}
for the cloud (supposing $L\gg\lambda_m$), where $\mu(0,L)=(\lambda_{m}/\lambda_{cx})^2\hbar\omega+c_{l,r}$, $G_*\equiv\sigma/\lambda_m$, and
\begin{subequations}\begin{align}
v_c(0)&=g^{\uparrow\downarrow}(\mu_l-\hbar\omega)/2\pi\hbar s,\\
v_c(L)&=g^{\uparrow\downarrow}(\hbar\omega-\mu_r)/2\pi\hbar s,
\end{align}\label{bcc}\end{subequations}
for the condensate. The reservoir spin accumulations are $\mu_l=\mu_r=0$ in the good spin sink case and are found according to $n_cv_c+j_x=0$ (at both interfaces) for the poor spin sinks. Integrating the steady-state version of Eq.~\eqref{lha},
\begin{equation}
\partial_xv_c=(\mu - \hbar\omega)/\tau_{cx}T -2\alpha\omega,
\label{dvc}
\end{equation}
we get for $\Delta v_c\equiv v_c(L)-v_c(0)$:
\begin{equation}
\Delta v_c=\frac{\lambda_m(c_l+c_r)}{\tau_{cx}T} -\left[2\alpha+\frac{\hbar(\lambda_m/\lambda_n)^2}{\tau_{cx}T}\right]\omega L.
\label{lvc}
\end{equation}

In the simpler, good spin sink case (where the spin Seebeck physics is manifested through the total spin currents injected into the metal reservoirs), we thus have 5 linear equations, \eqref{bcx}, \eqref{bcc}, and \eqref{lvc}, for 5 unknowns: $c_{l,r}$, $\hbar\omega$, and $v_c$ at $x=0,L$. For poor spin sinks (where the spin Seebeck physics is manifested through the spin accumulations induced in the metal reservoirs), we have two additional unknowns, $\mu_{l,r}$, and two more equations (for the vanishing total spin current at the interfaces). Note that the differential equation \eqref{lhb} for $\delta n_c$ decouples in the linearized treatment.
Adding and subtracting Eqs.~\eqref{bcx}, and substituting the difference of Eqs.~\eqref{bcc} into Eq.~\eqref{lvc} leads to
\begin{align}
(G+G_*)c_--\varsigma\Delta T/L-G \mu_{-}&=0,\nonumber\\
(G+G_*)c_++G(\lambda_m/\lambda_{cx})^2\hbar\omega- G \mu_{+}&=0,\nonumber\\
\frac{\lambda_mc_+}{\tau_{cx}T}  -\left[\alpha+\frac{\hbar(\lambda_m/\lambda_n)^2}{2\tau_{cx}T}+\frac{g^{\uparrow\downarrow}}{2\pi sL} \left( 1-\frac{\mu_{+}}{\hbar\omega } \right)\right]\omega L &=0,
\end{align}
where $c_\pm\equiv(c_l\pm c_r)/2$ and $\mu_\pm\equiv(\mu_l\pm \mu_r)/2$.

In the good spin sink case, $\mu_\pm=0$, the last two equations above lead immediately to $\omega=0$ and $c_+=0$. The remaining equation gives
\begin{equation}
c_l=\frac{\varsigma\Delta T/L}{G+G_*}=-c_r.
\end{equation}
The spin currents at the two interfaces (which turn out to be purely thermal and equivalent) are thus given by
\begin{align}
j_x=-\frac{\varsigma\Delta T/L}{1+G_*/G},
\label{jx}
\end{align}
and vanish when either $\lambda_{cx}\to0$ (strong condensate-cloud interaction regime, where $\lambda_m\to\lambda_{cx}$) or $\lambda_n\to0$ (strong magnon damping regime, where $\lambda_m\to\lambda_n$), since $G_*\propto1/\lambda_m\to\infty$. As, by decreasing field $B$, we go deeper into the condensate phase at a fixed $T$, and $n_c$ is monotonically increasing, $\lambda_{cx}$ decreases and thus the magnitude of $j_x$ is reduced (see Fig.~\ref{phasediagram}, where we took into account the dependence of $\lambda_{m}$ on $B$ but ignored the dependence of other quantities on $B$, which is valid as long as $T \gg K$  \cite{suppl1}). $j_x$ is largest at the transition point to the normal state and is given by Eq.~\eqref{jx} with $\lambda_m\to\lambda_n$.
Note that although the superfluid velocity $v_c$ vanishes at both interfaces, it is nonzero inside the ferromagnet (at distances beyond $\lambda_m$ from the interfaces), according to Eq.~\eqref{dvc}:
\begin{equation}
v_c=\frac{c_l\lambda_m}{\tau_{cx}T}=\frac{\varsigma\Delta T/L}{G+G_*}\frac{\lambda_m}{\tau_{cx}T}.
\end{equation}
Already in this simple case we encounter the \textit{conveyor-belt} physics, as the superfluid spin current $n_cv_c$ in the bulk counteracts the diffusive thermal flux $-\varsigma\Delta T/L$ and reduces the net spin Seebeck effect as measured at interfaces. 

\begin{figure}[t]
\includegraphics[width=0.9\linewidth]{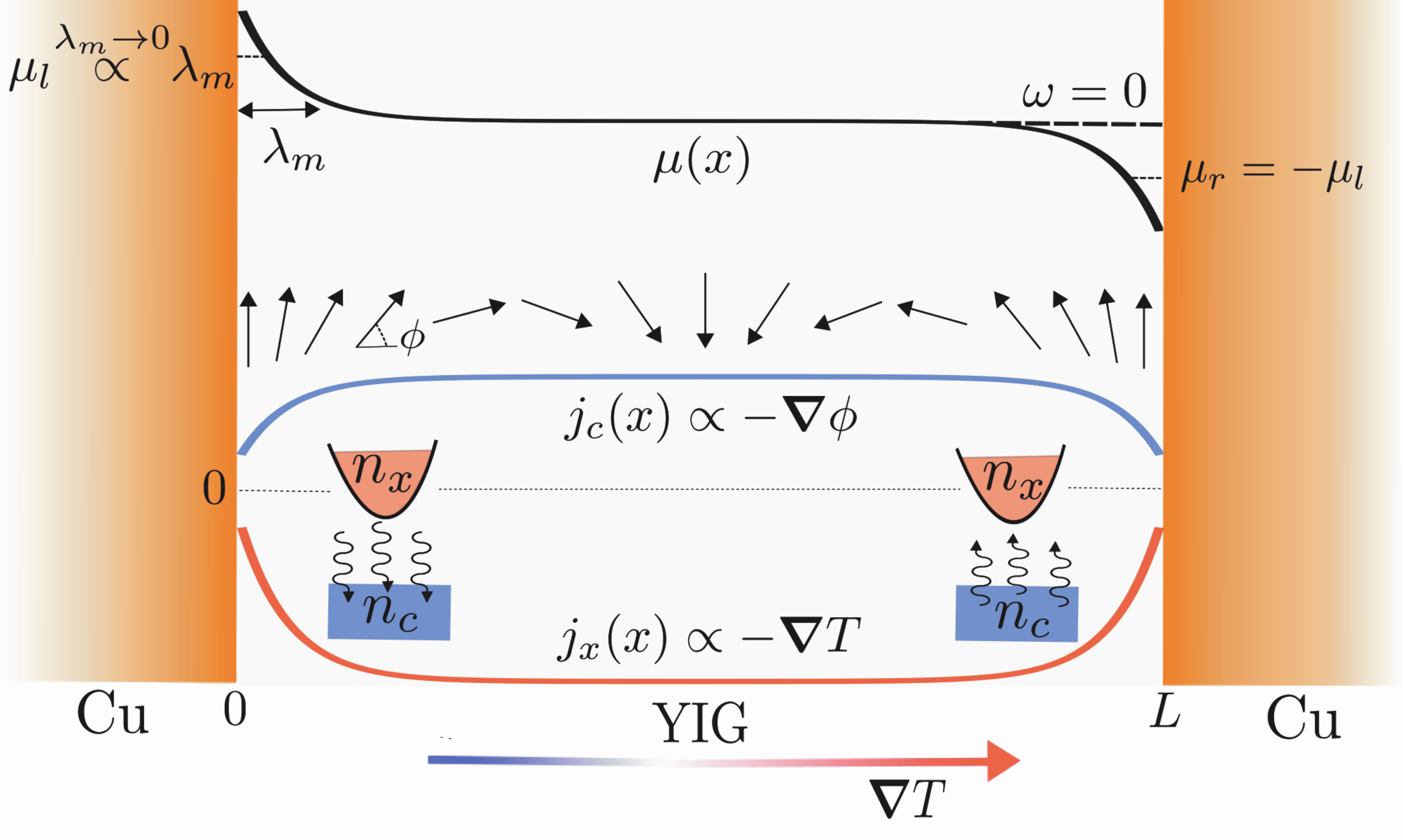}
\caption{In the presence of a temperature gradient $\Delta T$, the magnon chemical potential $\mu(x)$ deviates near the interfaces from its zero bulk value in the ferromagnet (YIG). This is accompanied by the electronic spin accumulation build-up in adjacent metals (Cu, treated as a poor spin sink). The spin accumulation $\mu_{l}$ at the left interface exerts a torque on the magnetic order parameter, twisting it in the opposite direction with respect to the one induced by $\mu_{r}=-\mu_{l}$ at the right interface. In the mirror-symmetric case, the precession frequency $\omega$ vanishes. The condensate, $j_c$, and thermal, $j_x$, contributions to the spin currents are plotted for $\lambda_{n}=\lambda_{cx}$.}
  \label{scheme3}
\end{figure}

In the opposite limit of the poor spin sinks, we still find $\omega=0$ and $c_+=0$, so that $\mu_{+}=0$, while
\begin{equation}
\mu_{l}= \frac{\varsigma\Delta T/L}{G_* + (1 + G_*/G)g^{\uparrow \downarrow} n_{c}/2\pi\hbar s}=-\mu_{r}.
\end{equation}
This spin accumulation vanishes when either $\lambda_{cx}\to0$ or $\lambda_n\to0$ and decreases with decreasing field $B$, displaying an analogous behavior to the one of the spin current at the interfaces in the good spin sink case (see Fig.~\ref{phasediagram}). While the total current now vanishes at the interfaces, $j_{x}$ and $v_{c}$ are both nonzero in the ferromagnet (see Fig.~\ref{scheme3}).

\textit{Discussion and conclusions.}|In this work, we constructed a hydrodynamic theory which describes the interactions between thermal and condensed magnons in an easy-plane magnetic  insulator in the presence of a thermal gradient. We predicted that spin superfluidity can be induced by sweeping the external magnetic field and experimentally probed via spin Seebeck effect. Although we have  explicitly considered  a ferromagnetic insulator, we anticipate, according to Refs.~\cite{Halperin} and \cite{arxiv}, qualitatively similar behavior also for antiferromagnets.  Future works should more systematically address the magnon-phonon relaxation mechanisms and study the role of magnons in the net heat transport. Nonlinear response, in the context of dynamic instabilities \cite{benderduine} and pinning by parasitic in-plane anisotropies \cite{sonin}, will be addressed elsewhere.

\acknowledgments
 
The authors thank R.~E. Troncoso and S.~Takei for helpful discussions.  This work is supported by the Stichting voor Fundamenteel Onderzoek der Materie (FOM) and is part of the D-ITP consortium, a program of the Netherlands Organisation
for Scientific Research (NWO) that is funded by
the Dutch Ministry of Education, Culture and Science
(OCW), and in part by the NSF-funded MRSEC under Grant No. DMR-1420451, and US DOE-BES under Award No. DE-SC0012190.

\end{document}


\title{Supplemental Material: Two-Fluid Theory for Spin Superfluidity in Magnetic Insulators}

\author{B. Flebus}
\affiliation{Institute for Theoretical Physics and Center for Extreme Matter and Emergent Phenomena, Utrecht University, Leuvenlaan 4, 3584 CE Utrecht, The Netherlands}
\author{S. A. Bender}
\author{Y. Tserkovnyak}
\affiliation{Department of Physics and Astronomy, University of California, Los Angeles, California 90095, USA}
\author{ R. A. Duine}
\affiliation{Institute for Theoretical Physics and Center for Extreme Matter and Emergent Phenomena, Utrecht University, Leuvenlaan 4, 3584 CE Utrecht, The Netherlands}

\maketitle
In this supplemental material, we derive the Boltzmann equation for the distribution density $f_{\textbf{p}}(\textbf{r},t)$ of the thermal magnons and establish the equilibrium state of the magnet, around which we expand for its linear response  to a temperature gradient.

\textit{Thermal cloud hydrodynamics}.|According to the Hamiltonian (1), the thermal magnons experience the Hartree-Fock mean-field potential $U=\hbar \Omega+ 2K n_c/s$, which lifts their parabolic spectrum. Note that it is split relative to the condensate frequency by $\Delta\equiv U-\hbar\omega=Kn_c/s$ [neglecting the inhomogeneous term in Eq.~(3b)] \cite{note}. At high temperatures, $T \gg U$ (setting $k_B\equiv1$, while still assuming that $T\ll T_c$, the Curie temperature), we can develop a mean-field transport theory for exchange magnons constituting the normal cloud, which undergo fast spin-preserving internal equilibration. The corresponding Boltzmann equation is given by
\begin{align}
\partial_{t} f +& \mathbf{p}\cdot\partial_\mathbf{r}f/m - \partial_{\mathbf{r}} U \cdot \partial_{\mathbf{p}} f \nonumber \\ 
  &=(f_p-f)/\tau_{\alpha} +  (\bar{f}-f)/\tau_{d}+C_{cx}+C_{xx},
\label{boltzmann} 
\end{align}
where $f_\mathbf{p}(\mathbf{r},t)$ is the Wigner transform of the field $\langle\hat{\phi}^\dagger(\mathbf{r'},t)\hat{\phi}(\mathbf{r''},t)\rangle$ [so that $n_x(\mathbf{r},t)=\int d^3pf_\mathbf{p}(\mathbf{r},t)/(2\pi\hbar)^3$], with $\bar{f}$ being the $\mathbf{p}$-space angular average of $f$ and $f_p$ is the phonon (Bose-Einstein) distribution function. The relaxation time $\tau_{\alpha}=\hbar/2\alpha (p^2/2m+U)$ describes the Gilbert damping (associated with the phonon bath), while the strength of elastic spin-preserving disorder scattering of magnons is parametrized by an energy-dependent time scale $\tau_{d}$. The scattering interactions with the condensate and among the thermal magnons are respectively described by the collision integrals $C_{cx}$ and $C_{xx}$ [17]. The latter scattering rate, which is governed by the exchange interactions, is expected to be fast at high temperatures [16], forcing the cloud towards a local Bose-Einstein profile with well-defined temperature $T$ and chemical potential $\mu$. The total magnon number conservation within the treatment of $C_{cx}$ based on Eq.~(1) dictates that $\Gamma_{cx}=\int d^3p \,C_{cx}/(2\pi\hbar)^3$.

The condensate-cloud scattering rate can be written as $\Gamma_{cx}=n_{c}(\hbar\omega-\mu)/\tau_{cx}T$ [17], where $\tau_{cx}$ is determined according to the Fermi's golden rule as
\begin{align}
\frac{1}{\tau_{cx}}=&\frac{2(K/s)^{2}}{(2\pi)^{5} \hbar^{7}} \int d^{3} p_{1} d^{3} p_{2} d^{3} p_{3} \; \delta (\mathbf{p}_{1} - \mathbf{p}_{2} -\mathbf{p}_{3})   \nonumber \\
& \times \delta ( \hbar \omega + \epsilon_{1} - \epsilon_{2} - \epsilon_{3} ) ( 1+ f_{1}) f_{2} f_{3},
\label{tcx}
\end{align}
where $\epsilon_i\equiv\mathbf{p}_i^{2}/2m + U$, $f_{i}\equiv f_B[(\epsilon_i-\mu)/T]$, and $f_{B}(x)\equiv ( e^x - 1)^{-1}$.  Eq.~(\ref{tcx}) can be evaluated in equilibrium ($U\to Kn_c/s$ and $\mu\to0$), in which case $\hbar/\tau_{cx}\propto(T/T_c)^3 K^2/T$, with a dimensionless prefactor that depends on $(K/T)(n_c/s)$. It is useful to remark that $\Gamma_{cx}\propto\hbar\omega-\mu$ is intuitively natural (as $\mu=\hbar\omega$ corresponds to the entropic extremum of the closed magnetic system).

The $g$ coefficients in Eqs.~(5) can be obtained according to the Gilbert-damping term $\propto\tau_\alpha^{-1}$ in Eq.~\eqref{boltzmann}, and go as $g_{n\mu}\sim  g_{nT} \sim \alpha s (T/T_c)^{3/2}/ \hbar$, $g_{uT} \sim g_{u\mu} = T g_{nT}$ (with the equality reflecting Onsager reciprocity). In practice, however, the magnon energy relaxation $\propto g_{uT}$ driven by the magnon-phonon temperature mismatch, $T-T_p$, can be dominated by nonrelativistic spin-preserving magnon-phonon scattering (which is clearly outside the Gilbert damping phenomenology).  We can similarly evaluate the coefficient appearing in Eqs.~(8) as \cite{bendertsesrkov} $G=\partial_{\mu} M_{1}$, $S=\partial_{T} M_{1}$, $K=\partial_{T} M_{2}$, and $
\Pi=\partial_{\mu} M_{2}$, in terms of
\begin{align}
M_{n}\equiv\frac{g^{\uparrow \downarrow}}{\pi\hbar s} \int \frac{d^3p}{(2\pi \hbar)^{3}} \epsilon_\mathbf{p}^{n}f_{B}[(\epsilon_\mathbf{p}-\mu)/T]. 
\end{align}
In equilibrium (in the condensate phase), $\epsilon_\mathbf{p}=p^2/2m + Kn_c/s$. Approximating, furthermore, $\epsilon_\mathbf{p}\approx p^2/2m$ (which is justified as $T\gg U\to Kn_c/s$), we get $(G,S,K,\Pi)=(g^{\uparrow\downarrow}/\pi\alpha s)(g_{n\mu},g_{nT},g_{uT},g_{u\mu})$, using the above Gilbert-damping expressions for the $g$'s.

The transport coefficients in Eq.~(6) are obtained following the integration of Eq.~\eqref{boltzmann} over momentum subsequent to multiplication by $\mathbf{p}$. They involve integrals over the magnon energies, with the integrands being proportional to the scattering time $\tau=1/(\tau_\alpha^{-1}+\tau_d^{-1})$. Assuming an energy-independent scattering length $l$, the scattering time $\tau=(2\alpha \epsilon_{\mathbf{p}}/\hbar+\sqrt{2\epsilon_{\mathbf{p}}/m}/l)^{-1}$ is dominated by disorder scattering at energies below $\epsilon^* \equiv T_c /s^{2/3}(\alpha l)^2$ and by Gilbert damping above $\epsilon^*$. With YIG in mind and taking $\alpha\sim 10^{-4}$, $l\sim\rm{\mu m}$,  and $s \sim 10/\rm{nm}^3$, one obtains $\epsilon^*> T_c$, in which case the transport coefficients are dominated by disorder scattering (which may be further supplemented by magnon-phonon scattering). Therefore, $ \varsigma  \sim (T/T_c)( s^{2/3} l)/\hbar$, $\kappa \sim \rho =  T \varsigma$, and $ \sigma  \sim(T/T_c)( s^{2/3} l)/\hbar$ (omitting a logarithmic factor for $\sigma$, which depends on the low-energy cutoff for our treatment of magnon transport). In practice, however, we may expect the transport coefficients to be further reduced due to magnon-phonon drag. It is important, for our purposes, to remark that the bulk spin Seebeck coefficient $\varsigma$ may in practice be strongly enhanced (compared to the above treament) by the magnon-phonon drag.

\textit{Equilibrium}.|Eqs.~(3b) and (\ref{boltzmann}) 
allow us to define the equilibrium conditions of the system, around which we will subsequently expand to determine its linear response to a thermal gradient. In a bulk equilibrium, all the quantities are spatially uniform, i.e., $\mathbf{v}_{c}=0$, and, due to Gilbert damping, both the condensate frequency $\omega$ (when $n_c\neq0$) and the magnon chemical potential $\mu$ must vanish (which is associated, in particular, with a vanishing scattering rate $\Gamma_{cx} \propto\hbar\omega- \mu$). The thermal cloud is then described by the Bose-Einstein profile $f_{B}[( p^2/2m+U)/T]$,  leading to the equilibrium density 
\begin{align}
n_x=\int\frac{d^3p}{(2\pi \hbar)^{3}} \frac{1}{e^{\frac{p^2/2m + K n_c/s}{T}} - 1} =\frac{1}{2}\left[\left(  1-\frac{B}{K} \right)s-n_c\right],
\label{density}
\end{align}
which provides the mean-field self-consistency relation for $n_c$. This allows to determine the transition temperature $T$ (as a function of external field $B<K$), where $n_c\to0$:
\begin{align}
T/T_c\to(1-B/K)^{2/3}/\Gamma_{3/2} \zeta_{3/2}.
\label{temperature}
\end{align}
Here $T_c\equiv s^{2/3} A$ estimates the Curie temperature of the magnet and $\Gamma$ and $\zeta$ are respectively gamma and Riemann zeta functions. For internal consistency, we need to verify the dilute Bose gas approximation, i.e., $n_c+n_x\ll s$, which restricts us to the vicinity of the critical point $(T/T_c,B/K)=(0,1)$ in the phase diagram, which is depicted in Fig. 2. (While the results at $B/K\sim0$ and/or $T/T_c\sim1$ are thus unreliable, we, nonetheless, still expect them to provide a good qualitative guidance.)
%